Nanoscale Minireview:

# Topological Crystalline Insulator Nanostructures

Jie Shen[1,2], Judy J. Cha[1,2]

[1] Department of Mechanical Engineering and Materials Science, Yale University, New Haven, CT, USA

[2] Energy Sciences Institute, Yale West Campus, West Haven, CT, USA

Topological crystalline insulators are topological insulators whose surface states are protected by the crystalline symmetry, instead of the time reversal symmetry. Similar to the first generation of three-dimensional topological insulators such as $Bi_2Se_3$ and $Bi_2Te_3$, topological crystalline insulators also possess surface states with exotic electronic properties such as spin-momentum locking and Dirac dispersion. Experimentally verified topological crystalline insulators to date are SnTe, $Pb_{1-x}Sn_xSe$, and $Pb_{1-x}Sn_xTe$. Because topological protection comes from the crystal symmetry, magnetic impurities or in-plane magnetic fields are not expected to open a gap in the surface states in topological crystalline insulators. Additionally, because they are cubic structure instead of layered structure, branched structures or strong coupling with other materials for large proximity effects are possible, which are difficult with layered $Bi_2Se_3$ and $Bi_2Te_3$. Thus, additional fundamental phenomena inaccessible in three-dimensional topological insulators can be pursued. In this review, topological crystalline insulator SnTe nanostructures will be discussed. For comparison, experimental results based on SnTe thin films will be covered. Surface state properties of topological crystalline insulators will be discussed briefly.

1. Introduction

As a new quantum matter, topological insulators (TIs) possess a unique electronic property; they exhibit gapped bulk bands with gapless surface or edge states that are massless and spin-polarized.[1] The surface or edge states are induced by the intrinsic bulk properties such as large spin orbit coupling, band inversion, and opposite parity of the bulk bands,[2] and topologically protected by symmetry of the material. Due to the topological protection, the surface states are immune to surface impurities or defects and cannot backscatter. A myriad of fundamental condensed matter physics phenomena as well as future electronic applications are expected in these materials.[3]

The TI field has grown dramatically since its first theoretical prediction by Kane and Mele in 2005 and the experimental verification in CdTe/HgTe quantum wells in 2007.[4] This burst of growth is largely due to the discovery in 2009 that traditionally well-known thermoelectric materials such as $Bi_2Te_3$ and $Bi_2Se_3$ turn out to be three-dimensional (3D) TIs.[5] A flurry of experiments were carried out to confirm the existence of the TI surface state in $Bi_2Te_3$ and $Bi_2Se_3$ and to study their electronic properties. These include direct imaging of the surface band structure by angle-resolved photoemission spectroscopy (ARPES),[6] linear band dispersion of the surface state and suppression of back scattering observed by scanning tunneling microscopy (STM),[7] transport verification of 2D Shubnikov-de Hass (SdH) oscillations which suggest a high mobility surface channel,[8] weak antilocalization indicating a mixture of surface states and bulk states[9] and transport measurement of Aharonov-Bohm (AB) oscillations in $Bi_2Se_3$ nanoribbons.[10] Efforts are ongoing to study the spin nature of the surface states in 3D TIs via transport and to study Majorana fermions by interfacing 3D TIs with superconductors.[11]

A new class of TIs, topological crystalline insulators (TCIs), was recently predicted and experimentally verified, expanding the ever-growing TI materials family.[12] Experimentally confirmed TCIs are SnTe, $Pb_{1-x}Sn_xSe$, and $Pb_{1-x}Sn_xTe$.[12a, 12c, 13] For $Pb_{1-x}Sn_xSe$, and $Pb_{1-x}Sn_xTe$, a transition from a trivial insulator to a TCI occurs at a critical concentration, $x_c$. For example, $Pb_{1-x}Sn_xSe$ becomes a TCI at x=0.23.[13a] TCIs are similar to 3D TIs in that they possess spin-polarized, linearly dispersive surface states. However, there are major differences between TCIs and TIs. TCIs possess multiple surface states specific to particular crystal surfaces while 3D TIs generally have a single surface state (BiSb alloy is an exception with five surface states[14]). Topological protection for TCIs comes from the crystal symmetry instead of the time reversal symmetry, which is the case for 3D TIs. These differences open up new opportunities to pursue fundamental phenomena that may not be possible with 3D TIs.

In this mini-review, nanostructured TCI SnTe will be discussed. For 3D TIs and TCIs, nanostructuring is advantageous because the surface state can be enhanced dramatically due to a much higher surface to volume ratio of nanostructures compared to the bulk.[15] Before discussing SnTe nanostructures, the theory of topological crystalline insulators and experiments based on TCI bulks and thin films will be briefly reviewed. Synthesis and transport measurements of SnTe nanostructures will be covered in detail. Distinct advantages specific to nanostructures for studying topological surface states will be highlighted. The review will conclude with ongoing efforts to improve the TCI materials and future studies based on TCI SnTe materials.

## 2. Topological Crystalline Insulators

Unlike the three known 3D TIs - $Bi_2Se_3$, $Bi_2Te_3$, and $Sb_2Te_3$, which are layered structure,[5] TCIs have a rock salt crystal structure (Figure 1A).[12a, 12b] Crystal symmetry present in specific crystal surfaces such as {111} and {100} provides topological protection for surface states in TCIs.[12a] This is the main difference between TCIs and 3D TIs whose surface states are protected by the time reversal symmetry. Also, unlike $Bi_2Te_3$ and $Bi_2Se_3$ that possess a single surface state,[5] TCIs possess multiple surface states.[16] For SnTe, four surface states are present on each of {100}, {111}, and {110} surfaces.[16] Figure 1A shows surface states on {111} and {001} surfaces. Band structures of these surface states depend on the crystal symmetry.[16-17] For surface states on {100} surfaces, two surface states are located off of X, merging at high energies away from the Dirac point. On {111} surfaces, one surface state is centered at Γ and three surfaces are at M. Figure 1B shows the band structure calculation of SnTe, showing the surface state band along the Γ to M direction.[12a]

## 3. Experimental Results From SnTe Bulk and Thin Films

SnTe is a narrow-gap p-type semiconductor. Due to Sn vacancies, it is heavily doped with carrier densities ranging between high $10^{19}$ to $10^{21}$ cm$^{-3}$.[18] This is similar to 3D TI Bi$_2$Se$_3$ where Se vacancies lead to high bulk carriers ranging between high $10^{18}$ to $10^{21}$ cm$^{-3}$.[8b, 19] Similar to 3D TI Bi$_2$Te$_3$ and Bi$_2$Se$_3$, which are well-known thermoelectric materials,[20] SnTe also shows a good thermoelectric property and has been extensively studied.[21] For example, nanostructuring, texturing, doping, and alloying of SnTe were carried out to improve the thermoelectric property.[21b, 22] In addition to excellent thermoelectric properties, SnTe is known to undergo structural phase transitions. The first phase transition is from its room-temperature cubic structure to low-temperature rhombohedral structure.[23] This structural phase transition induces a ferroelectric phase transition.[23a] The transition shows up as a kink in resistivity as a function of temperature. The phase transition temperature scales with the bulk carrier density; it decreases with increasing carrier density.[23b] The second phase transition is from the rhombohedral structure to orthorhombic at even lower temperatures.[24] This second phase transition induces an antiferroelectric phase transition. Room-temperature ferromagnetism has been demonstrated in magnetically doped SnTe, doped with 3$d$ transition metals such as Cr and V.[25]

With renewed interest as TCIs, SnTe and its alloys, such as Pb$_{1-x}$Sn$_x$Se, and Pb$_{1-x}$Sn$_x$Te, have been synthesized as single-crystal bulks, thin films, and nanostructures. For thin films, SnTe and its alloys were grown on several substrates such as Bi$_2$Te$_3$, SrTiO$_3$(001), and Si(111) via molecular beam epitaxy (MBE).[26] Topological surface states have been verified by ARPES, which showed linearly dispersive two-dimensional bands, in agreement with theoretical calculations.[12c, 13b, 17, 27] Band structures of surface states on

{100} and {111} surfaces have been separately investigated to confirm the differences in surface state band structures based on crystalline symmetry.[16] The phase transition from a trivial insulator to a topological crystalline insulator was studied by APRES and Landau level spectroscopy using scanning tunneling microscopy and spectroscopy in $Pb_{1-x}Sn_xSe$[13a, 28] and $Pb_{1-x}Sn_xTe$.[13b, 29]

Due to the high bulk carrier densities induced by Sn vacancies, measuring the properties of surface states via transport is difficult with bulk SnTe. In the case of 3D TI $Bi_2Se_3$ and $Bi_2Te_3$, thin flakes can be obtained from the bulk by mechanical exfoliation exploiting the layered crystal structure and weak van der Waals interactions between the layers.[30] In the case of cubic SnTe, this is difficult.[31] For this reason, bulk transport studies on SnTe are few. Thin films or nanostructures provide more suitable platforms for transport measurements of the surface states due to higher surface to volume ratios. 2D SdH oscillations were observed in SnTe thin films grown on a buffer $Bi_2Te_3$ layer,[26c] as shown in Figure 2. Figure 2A and 2B show atomically smooth $Bi_2Te_3$ and SnTe before and after the growth respectively, indicating that $Bi_2Te_3$ can serve as a growth substrate to provide opportunities to couple surface states of different topological insulators. Figure 2C and 2D show SdH oscillations observed in the SnTe films. Angle-dependent analysis indicates that the SdH oscillations are 2D in nature. From the oscillation frequencies, surface carrier densities were estimated to be 2.6 and $3.4 \times 10^{11}$ cm$^{-2}$. The estimated surface carrier densities of SnTe are similar to those of 3D TI $Bi_2Te_2Se$ and $Bi_2Se_3$ nanoribbons coated with a ZnO coating layer.[32] The surface carrier mobility of SnTe was estimated to be 2000 cm$^2$/Vs from the observed SdH oscillations(the reference?). In comparison, the surface carrier mobility of 3D TI $Bi_2Te_2Se$ was estimated to be 1450 cm$^2$/Vs.[32a] The landau level plot from gate-dependent SdH oscillations shows an intercept of 0.55, indicating a Berry phase of $\pi$ (Figure 2D). This is due to the Dirac dispersion of the surface state. Weak anti-localization effects were also observed and analyzed in thin SnTe films grown on $BaF_2$ (001).[33] The weak anti-localization feature was fitted to 2D Hikami-Larkin-Nagaoka equation where the parameter, $\alpha$, denotes the number of conduction channels.[34] One topological surface state would result in $\alpha$ of 0.5.

The values for α ranged between 0.75 and 2.5. It was concluded that up to 5 quantum coherent weak antilocalization channels exist for SnTe, which is possible given multiple surface states that may contribute to the transport. However, due to possible coupling between the bulk and the surface and between the surface states, it is difficult to use weak antilocalization to study the surface states.

## 4. SnTe nanostructures

### 4.1. Benefits

SnTe nanostructures offer several distinct advantages compared to the thin film geometry for studying the surface states. First, besides the thin film growth, direct nanostructure growth is the only other way to increase the surface to volume ratio to enhance surface state effects. For $Bi_2Se_3$ and $Bi_2Te_3$ TIs, mechanical or chemical exfoliation is possible to obtain thin flakes from bulk to maximize the surface states because these are layered materials.[35] SnTe is a cubic structure, thus mechanical cleaving is difficult. Second, with SnTe nanostructures grown by vapor-liquid-solid (VLS) and vapor-solid (VS) growth methods, there is no strain in SnTe due to lattice mismatch between SnTe and the growth substrate. This is because VLS-grown SnTe nanostructures shoot off from the substrate, instead of laying flat on the substrate. For thin films grown by MBE, lattice matching is critical at the interface between SnTe and the substrate to preserve the high transport quality of the surface states. Third, multiple surface states of different band structures can be exposed in SnTe nanostructures by controlling the nanostructure morphology with changes in growth conditions. In thin films, only one crystal surface is exposed as a top surface. In contrast, nanostructures can contain multiple faceted surfaces. SnTe nanowires, nanoplates, and nanocubes have been grown with many {111} and {100} exposed surfaces.[36] Lastly, the nanostructure size is often comparable to coherent lengths of electrons, thus naturally enables study of the quantum nature of the surface states. In particular, nanostructures possess well-defined cross-sections, which are suitable for transport experiments. For example, the rectangular cross section of a SnTe nanowire combined with the nanoscale sets up an ideal situation for interference type transport measurements such as AB oscillations.[36d]

**4.2. SnTe nanostructure growth**

SnTe nanostructures can be grown by both wet and dry methods. SnTe nanoparticles and nanorods were previously grown using chemical bath and solvothermal methods where their morphology was controlled by growth conditions and different ligands.[37] For careful transport measurements of the TCI surface states however, high quality single crystalline nanostructures with clean surface termination are desired. For this purpose, VLS and VS growth methods have been used to grow single SnTe nanostructures.[36] Despite the relatively low vacuum and less control during growth compared to epitaxial growth methods like MBE, the crystalline quality of SnTe nanostructure is high. Transmission electron microscopy analysis shows single crystalline SnTe nanoribbons and nanoplates without any obvious defects.[36a, 38] The high quality crystallinity is supported by transport measurements, which show comparable bulk carrier densities to thin films and show coherent oscillations such as SdH oscillations in SnTe nanowires.[36d]

Figure 3 shows SnTe nanostructures with various morphologies grown via VLS and VS growth modes. For VLS growth, Au nanoparticles or Au thin films were used as growth catalyst. Nanowires and nanoplates have been observed. The cross section of the nanowires is rectangular, reflecting the cubic crystal structure of SnTe, and is determined by the original size of the Au nanoparticles.[36b] Sometimes, SnTe nanowires with a decreasing diameter are observed[36a]. SnTe nanoplates are somewhat unexpected as SnTe is not a layered crystal. The key factor in obtaining the nanoplate form is short growth time and lower substrate temperature.[36a] Because of surface energy differences, generally only {100} and {111} surfaces are exposed in these nanostructures.[36b] Indeed, for SnTe nanoplates, the top surface is observed to be either {100} or {111}. SnTe nanowires can either be smooth, grown along <100> direction or zigzagged. For {111} surfaces, surface energy calculation shows that they should be Te-terminated rather than Sn-

terminated.[36b] Without Au catalysts, SnTe nanocubes are grown, reflecting the underlying rock-salt crystal structure.

**4.3 SnTe nanostructure transport**

A couple of studies report transport measurements on SnTe nanostructures.[36a, 36d] Similar to SnTe bulk and thin films, the carrier density of SnTe nanostructures is high ranging between $10^{19}$ to $10^{21}$ cm$^{-3}$.[36a] Despite the high bulk carrier density, the unique square cross section of the SnTe nanowires provides an opportunity to probe surface states. Similar to AB oscillation studies in $Bi_2Se_3$ nanoribbons,[10, 39] SnTe nanowires also show pronounced AB oscillations, which point to presence of surface carriers (Figure 4B).[36d] SdH oscillations were also observed in these nanowires.[36d] A landau level fan diagram was constructed from the SdH oscillations to show a non-zero intercept of 0.42. This is in agreement with the SnTe thin film study[26c] and again indicates a presence of a surface state with linear dispersion. Transport measurements on SnTe nanoplates show a structural phase transition from the rock-salt to rhombohedral structure, indicated by a kink in resistivity around 47 K as shown in Figure 4A.[36a] At a very high carrier density, electron-electron interactions are also observed[36a]. Interestingly, weak antilocalization was not observed in these SnTe nanoplates. This may be due to the fact that the nanoplates were much thicker, ranging between ~ 50 nm to over 100 nm, than thin films that showed weak antilocalization.

**5. Materials Perspectives and Future studies**

Despite the initial successful experimental results on SnTe as a TCI such as direct observation of multiple surface states by APRES and 2D SdH and AB oscillations via transport, much work is left to do. Many of the challenges for SnTe TCIs are similar to those for $Bi_2Se_3$ and $Bi_2Te_3$ 3D TIs. Most immediately, the bulk carrier density needs to be reduced in order to study the surface state clearly. For bulk and thin film growths, Bi and Pb have been used to decrease the bulk carrier density.[13b, 26a, 27b, 40] For nanostructure growth, doping can sometimes be much more challenging than for bulk growth.

Effective charge doping to reduce the bulk carrier density while maintaining high mobility for the surface states is the most immediate step. Compensation charge doping using Bi or Pb powder during the nanostructure synthesis of SnTe can be tried. In the case of 3D TI $Bi_2Se_3$ nanoribbons, Sb was successfully used to reduce the bulk carrier density dramatically.[32b] Also, excess Sn vapor can be introduced during the SnTe nanostructure growth to minimize Sn vacancies. In addition, surface oxidation needs to be prevented. Similar to $Bi_2Te_3$ and $Bi_2Se_3$ which show degradation of surface states over time due to sample aging and surface oxidation,[41] the surface states of SnTe seem to degrade over time.[26c] *In situ* coating of SnTe with an insulating protective layer is highly desired.[42]

Currently, efforts are being made to make topological superconductors.[1b, 43] For $Bi_2Se_3$ 3D TIs, Cu intercalation was used to demonstrate superconductivity in $Bi_2Se_3$.[44] For SnTe, In doping leads to superconductivity where the superconductivity transition temperature increases with the In doping concentration.[45] The transition temperature is ~ 1.3 K for ~ 5% In doping.[45b] Indium doping has not been demonstrated in SnTe nanostructures. Given the unique transport measurements possible with SnTe nanowires but not with the bulk or thin films, this should be pursued. Magnetic doping in SnTe will also be interesting. Unlike 3D TIs whose surface states will be gapped by magnetic impurities because they are topologically protected by time reversal symmetry,[46] surface states of TCIs will not be gapped by magnetic impurities because the topological protection comes from the crystalline symmetry[12a]. It is predicted that depending on the direction of magnetic fields, surface states on {100} of TCIs can remain gapless or can be gapped, leading to quantum anomalous Hall effect.[47] Particularly, it will be interesting to study the coupling between ferromagnetism and the spin texture of the surface states. How the surface states are modified by the structural distortion due to the phase transition from rock salt to rhombohedral needs to be characterized carefully. The phase transition is predicted to open a small gap in the surface states on {100} as the crystal symmetry will be lost after the transition.[12c, 47] Strain-induced or pressure-induced band gap engineering is being predicted for SnTe TCIs.[48] Applying strain or pressure to other rock salt crystals can also induce a topological phase transition.[49]

SnTe may serve as an ideal platform for Majorana fermion braiding operations because of its cubic structure. To couple Majorana fermions, junction structures are required.[50] Such structures will be difficult to obtain with $Bi_2Se_3$ and $Bi_2Te_3$ as they are layered structures. With SnTe, exploiting the rock-salt crystal structure, it is possible to synthesize branched nanostructures. Branched nanostructures have been demonstrated for PbSe,[51] which is also cubic. First, careful Majorana fermion studies should be carried out by putting superconducting contacts to SnTe. These studies can be compared to similar studies using $Bi_2Se_3$ and $Bi_2Te_3$.[11b, 30c, 52] Interfacing different topological insulators by constructing heterostructures such as SnTe on $Bi_2Se_3$ or $Bi_2Te_3$ may be interesting to induce exotic coupling between topological surface states with different band structures. Finally, unambiguous demonstration of the spin polarization (spin-momentum locking property) needs to be achieved for SnTe.[27a, 53] For $Bi_2Se_3$, it was demonstrated by observing switch of the photo-current direction when a $Bi_2Se_3$ thin flake was excited with clockwise and counter-clockwise circularly polarized light.[11a]

## 6. Conclusion

SnTe has been studied previously as a lead-free thermoelectric material. Recently, it was rediscovered as a topological crystalline insulator, exhibiting multiple surface states that are spin-polarized and linearly dispersive. The surface states are topologically protected by the crystal symmetry of SnTe, thus they are robust against surface impurities. The exotic nature of the topological surface states open opportunities to study fundamental condense matter physics such as Majorana fermions and spin quantum hall effect. Nanostructured SnTe is particularly promising because the surface states will be enhanced due to the high surface to volume ratio. Already, SnTe nanowires and nanoplates have been demonstrated with transport measurements indicating the presence of high-mobility surface states. For future studies, bulk carrier densities need to be reduced and the surface of SnTe should be protected by a coating layer. Branched SnTe structures will be particularly interesting for coupling surface states and braiding Majorana fermions.

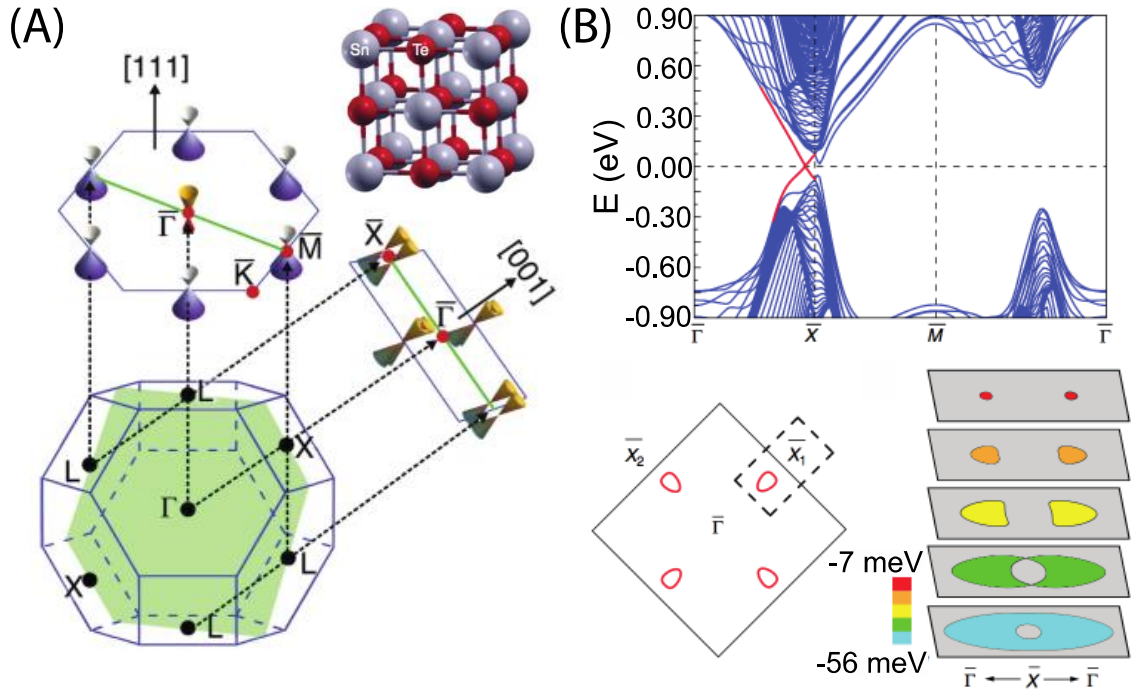

**Figure 1.** Atomic and electronic band structure of a topological crystalline insulator, SnTe. (A) The crystal structure of SnTe and the face-centered-cubic Brillouin zone. Surface states are present on {100}, {111}, and {110} surfaces. (B) Band structure calculation of the {001} surface of SnTe. The surface band is denoted in red. Bottom left shows the Fermi surface while the bottom right shows Fermi surfaces at different energies, exhibiting a Lifshitz transition. Reprinted with permission from ref. 12*a* and 17.

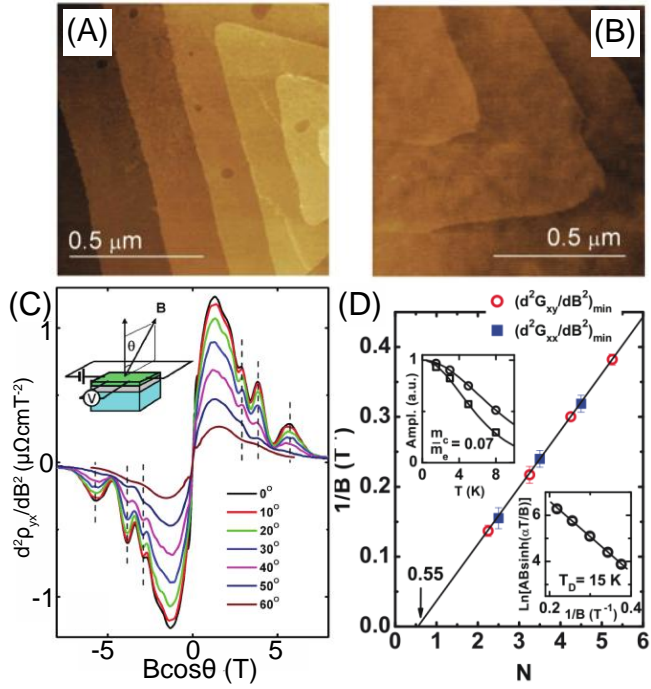

**Figure 2**. SnTe thin film grown on $Bi_2Te_3$ by MBE. (A) Atomic force microscope (AFM) image of the $Bi_2Te_3$ layer showing atomically flat terraces. (B) AFM image of the SnTe film grown on the $Bi_2Te_3$ layer. The film is atomically smooth. (C) Second derivative of the Hall resistances at various angles are plotted as a function of the perpendicular component of the magnetic field. The observed SdH oscillations align on top of each other, indicating the 2D nature. (D) Landau level index plot shows a non-zero intercept of 0.55, indicating the $\pi$ Berry phase. Reprinted with permission from ref. 25*c*.

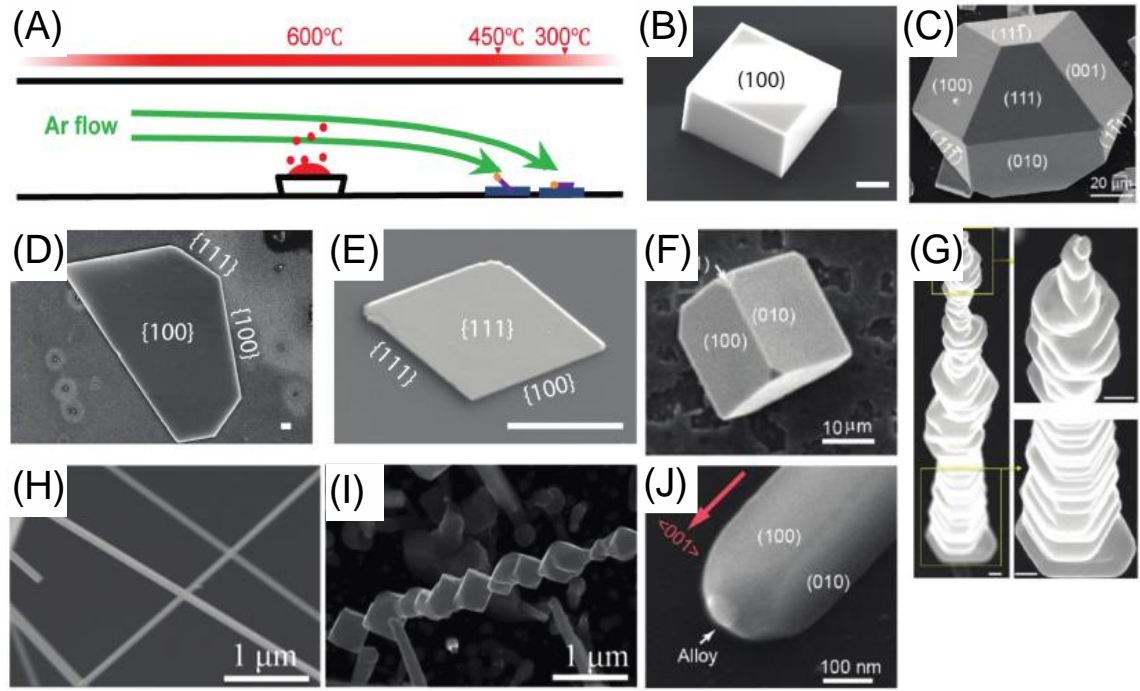

**Figure 3**. SnTe nanostructures via VLS and VS growth methods. (A) Growth schematic where SnTe source powder is heated to a growth temperature at the center of the furnace. Growth substrates are placed at the colder zone to collect SnTe nanostructures. (B) SnTe nanoblock without Au catalyst. (C) SnTe nanostructure without Au catalyst in Te rich environment. (D, E) SnTe nanoplates with {100} and {111} as top surfaces, respectively. (F) SnTe nanocube without Au catalyst in Te poor environment. (G) Zigzag SnTe nanowires. (H) SnTe nanoribbons with smooth surfaces. (I) Zigzag SnTe nanowire. (J) SnTe nanowire with {100} surfaces. Scale bars in B, D, and E are 2 μm. Scale bars in G are 200 nm. Reprinted with permission from ref. 34*a*, 34*b* and 36.

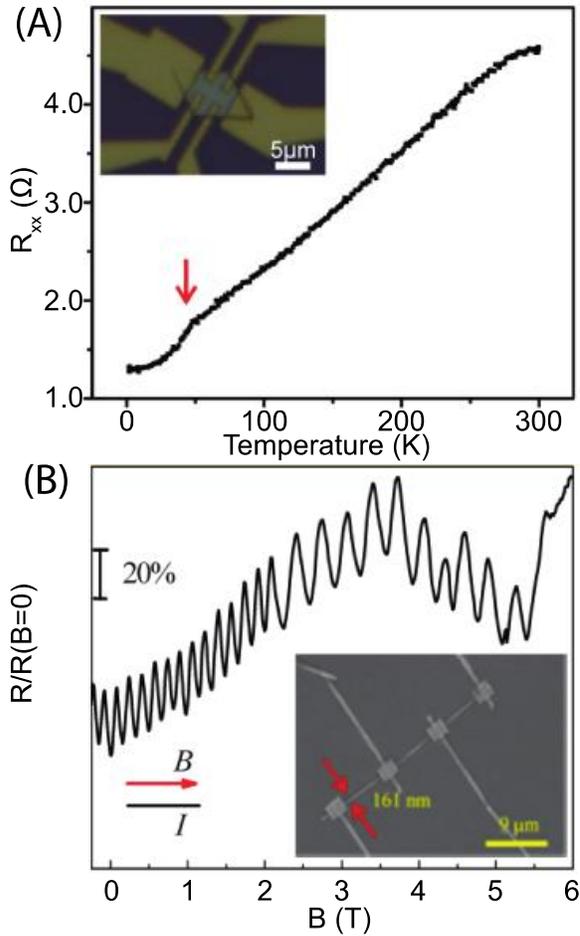

**Figure 4**. Transport on SnTe nanostructures. (A) $R_{xx}$ versus T curve of a (111) nanoplate shows a kink at ~ 47K, indicating a structural phase transition. Inset shows the measured device. (B) Normalized magnetoresistance of a narrow SnTe nanowire with parallel magnetic fields at 20 K. Oscillations with two frequencies, *h/e* and *h/2e*, are observed, which are Altshuler-Aronov-Spival and AB oscillations. Inset shows the device. Reprinted with permission from ref. 34*a* and 34*d*.